# M-BEHZAD: Minimum distance Based Energy efficiency using Hemisphere Zoning with Advanced Divide-and-Rule Scheme for Wireless Sensor Networks

Muzammil Behzad, *Student Member*, IEEE

*Abstract* — Routing Protocols are engaged in a vigorous fashion to boost up energy efficiency in WSNs. In this paper, we propose a novel routing protocol; Minimum distance Based Energy efficiency using Hemisphere Zoning with Advanced Divide-and-Rule scheme (M-BEHZAD), to maximize network lifespan, throughput and stability period of the sensors deployed in an un-attended network zone. To accomplish these objectives, static clustering technique along with threshold conscious transmissions have been used. The robustness of our proposed scheme lies in its Cluster Heads (CHs) selection and network field division which we are introducing as 'Hemisphere Zoning (HZ)'. We have implemented 3-Tier architecture to minimize the communication distance which not only leads to a better network performance but also significantly reduces the energy and coverage holes, and results in a longer stability period. We have also utilized Uniform Random Model (URM) to compute packets dropped to make our scheme a more practical approach. Results from comprehensive simulations using MATLAB validate its applicability.

## I. INTRODUCTION

Wireless Sensor Networks (WSNs) consist of randomly deployed and limited energy nodes that are capable of sensing different attributes in real time and can provide efficient and reliable communication using a microwave link. These nodes sense data and forward it to Base Station (BS) using direct and/or multi-hop communication. BS then processes and forwards it to the end users. WSNs have different type of nodes such as acoustic, underwater, seismic, visual, radar, thermal and many others. Their application includes, but not limited to, military, healthcare, environmental and commercial.

Optimization of energy consumption to elongate network lifetime has been one of the hot research topics in WSNs. To address this problem, much of the work has been done during recent years where efficient utilization of energy, by proposing robust protocols, was the main target. Clustering is of the technique that's used to avoid inefficient use of energy. Clustering may be dynamic or static. In dynamic clustering, cluster and number of nodes associated with cluster vary. In static clustering, cluster and number of nodes associated with cluster remain fixed throughout the network lifespan.

WSN may be either proactive or reactive based on its nature. In the former case, nodes periodically switch on their sensors and transmitters, sense the required attribute from environment and transmit it. Thus, these are well suited for periodic data monitoring applications. In the latter case1, the sensors respond instantaneously to abrupt or extreme changes in a specific attribute and keep their transmitters off otherwise. Hence, these are suitable for time critical applications.

Generally, sensors in WSNs are equipped with some amount of initial energies. Based on this amount, WSNs may be homogenous or heterogeneous. In homogenous networks, all the nodes deployed in network field have same initial energies. In heterogeneous network, the initial energies of some nodes are different by a certain factor and as a result those are called advanced nodes. Heterogeneous sensor networks may be two-level or multilevel.

In this paper, we propose and evaluate a new reactive and less energy hunger clustering protocol; M-BEHZAD: Minimum distance Based Energy efficiency using Hemisphere Zoning with Advanced Divide-and-Rule scheme for heterogeneous WSNs. We utilize reactive as well as heterogeneous characteristics of the wireless sensor network to ultimately prolong network lifetime. Additionally, Hemisphere Zoning (HZ) is introduced for network field division.

Rest of the paper is organized as follows: section II presents the related work, while motivation is presented briefly in Section III. Section IV deals with detailed implementation of our proposed approach, i.e. M-BEHZAD. Simulations, results and discussions are explained in Section V, and finally section VI concludes the paper.

## II. RELATED WORK

Cluster based routing protocol, LEACH [1], is proposed by W. Heinzelman. Multi-hop communication scheme is used in this protocol. Selection of CHs is probabilistic; therefore, distribution of CHs is not uniform which result in unbalanced distribution of CHs in the network. A. Manjeshwar, *et al.*, [2] proposed a reactive protocol, TEEN. This protocol outperforms the traditional schemes in terms of lifetime. However, it is limited for temperature specific applications.

Authors of [3] and [4] introduced heterogeneous wireless sensor networks with different amount of node initial energies. In [3], the authors proposed SEP scheme for two-level heterogeneous WSN, which mainly consists of two different type of nodes based on initial energy. The advanced nodes are equipped with more initial energy than the normal nodes. SEP prolongs the stability period. However, it is not appropriate for widely used multi-level heterogeneous wireless sensor networks, which include more than two types of nodes. In [4], DEEC protocol is proposed which take advantage of multi-level heterogeneity. CHs selection in DEEC is based on high initial and residual energy of nodes. Hence, DEEC achieves longer lifetime than SEP.

In [5] and [6], static clustering technique has been implemented. Authors in [5] proposed a Regional Energy Efficient Cluster Heads based on Maximum Energy (REECH-ME) by using static clustering technique. Selection of CHs is on the basis of residual energy. The protocol achieves longer network lifetime than LEACH [1]. However, energy holes may

*Correspondance\*: Muzammil Behzad, muzammil.behzad@ieee.org, http://muzammilbehzad.com*

be created due to unequal areas of static clusters. Therefore, both fail to uniformly distribute the nodes in divided regions.

A. Ahmad, *et al.*, [7] introduced a new routing technique, Density Controlled Divide-and-Rule (DDR) for Wireless Sensor Networks. They solve the problem of unbalanced energy utilization that causes energy holes and coverage holes in WSNs. A hybrid approach of uniform-random deployment of nodes is used in this work. The protocol beats LEACH [1] and REECH-ME [5] in terms of energy consumption.

Authors in [8] presented a distributed coverage hole repair algorithm (HORA) for WSNs and discussed about the energy hole problem due to frequent sensing and non-uniform nodes distribution. They proposed a pixel-based transmission scheme to mitigate inefficient use of energy. Energy hole is created due to non-uniform distribution of the nodes. In hot-spots regions, energy is consumed at very fast rate, due to which nodes behind the energy hole cannot transmit to BS. In order to repair coverage holes of network, nodes with high degree of density are moved meanwhile keeping the coverage degree of the neighbors of that moving node same. It's observed that substantial amount of coverage overlapping can be minimized and percentage of coverage of the holes can be maximized. The paper overall succeeded to achieve the targets.

### III. MOTIVATION

In this section, we elaborate the flaws in traditional routing protocols. Energy is a very scarce resource and must be used very efficiently, especially when it's the case of portable and non-rechargeable power supply. During the recent years, much research has been done in the area of wireless sensor networks. However, lack of significant attention has been noted that is being given to the time critically of applications. Most of the existing protocols give attention to reduce coverage holes and energy holes to maximize network life time but they assume a sensor node to collect and transmit a sensed attribute periodically from its environment. Also, these protocols select the CHs randomly and as a result cannot perform up to the mark. Therefore, the selection criterion can be improved.

We believe that there exists a need for the sensor networks to have an improved CHs election mechanism and to be reactive, i.e., respond only to drastic changes or a sudden event. We also believe that the dynamic clustering technique in the existing reactive protocol, TEEN [2], leads to quick death of the network. Additionally, heterogeneity in sensor networks prolongs the network lifetime. So, to overcome the aforementioned issues, there exists a need to design new routing protocols for WSNs.

### IV. THE M-BEHZAD SYSTEM DESIGN

Firstly, radio model of the subject field is presented. Then we describe Hemisphere Zoning (HZ) based network model, and then 3-Tier communication architecture in these zones is explained. Afterwards, the CHs selection mechanism and protocol operation is described.

#### A. Radio Model

In our work, we have assumed a simple first order radio model for transmitting information from one node to other as shown in fig. 1. The radio model uses $E_{elec}=50\ nJ/bit$ for powering the transmitter and/or receiver circuitry. It uses free space energy of $\varepsilon_{fs} = 10\ pJ/bit/4m^2$ and multipath energy of $\varepsilon_{mp} = 0.0013\ pJ/bit/m^4$. CHs consume $E_{DA}=50\ nJ/bit$ to aggregate the data received from normal nodes. The radio parameters used in our work are shown in table 1.

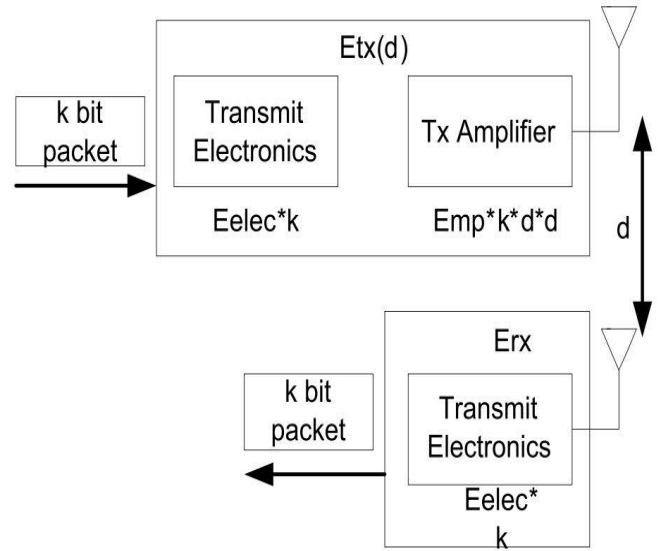

Fig. 1 Radio Model

We also take care of $d^2$ energy losses due to transmission over a non-ideal channel. Accordingly, to transmit a k-bit message on distance d, the mathematical expressions are:

$$d_o = \sqrt{\frac{\varepsilon_{fs}}{\varepsilon_{mp}}} \qquad (1)$$

$$\text{if } d < d_0 \qquad E_{T_x}(k,d) = E_{elec}*k + \varepsilon_{fs}*k*d^2 \qquad (2)$$

$$\text{if } d \geq d_0 \qquad E_{T_x}(k,d) = E_{elec}*k + \varepsilon_{mp}*k*d^4 \qquad (3)$$

$$\text{Reception Energy:} \qquad E_{R_x}(k) = E_{elec}*k \qquad (4)$$

#### B. Hemisphere Zoning (HZ) Based Network Model

We have taken *100m* x *100m* area for the network field with BS at central reference point i.e. $C_p(x_1, y_1)$, where the area has been divided by introducing the concept of HZ. The field is associated with earth's geographical form and is taken as a sphere of certain radius *ηβ*.

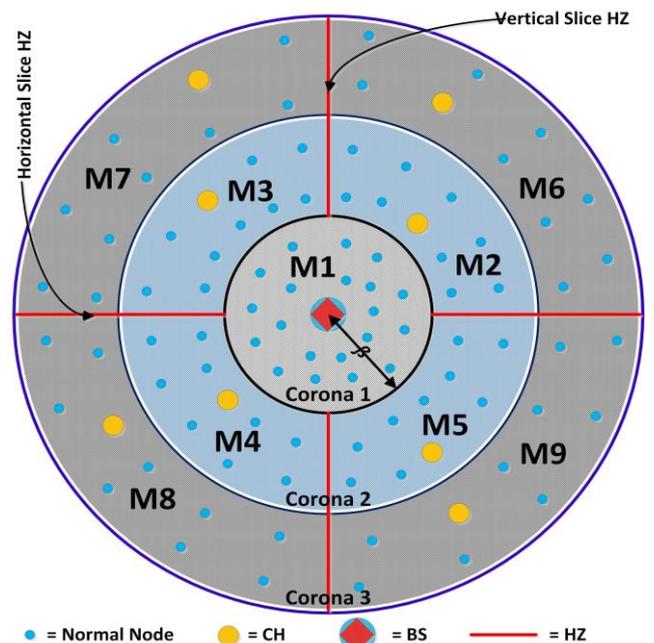

Fig. 2. HZ based Network Model and Nodes Deployment

From the concepts of geometry, the observation area is then divided into hemispheres firstly by slicing the sphere horizontally and then vertically. This results in Vertical Slice HZ and Horizontal Slice HZ. After that, the field is divided it into $\eta$ coronas with center of each as coordinates of BS (In our case, we took $\eta=3$). A total of 100 nodes are deployed uniformly in the field. The radius of the innermost, middle and outer circles is $\beta$ m, $2\beta$ m and $3\beta$ m, respectively. Consequently, 9 regions are formed in this way. The nodes are divided equally in 9 regions so every region gets 10 nodes fixed in every round except region 1 which gets 20 nodes. The nodes will be deployed in their corresponding regions randomly which results into uniform-random deployment as already stated in the previous sections. Separating the network into zones assists in minimizing the distance between cluster members and CHs. This is depicted in fig 2. As we have used static clustering, the number of clusters and the CHs is fixed.

*1) Cluster Formation:* The entire network field is divided into $\eta=3$ equidistant coronas and into nine regions. Nodes are deployed randomly and distributed uniformly in each cluster. The value of $\eta$ depends upon network field area and number of deployed nodes in order to reduce the communication distance between nodes, CH and base station. Hence, value of $\eta$ is set to 3 in our work. The coronas are thus named as: Internal Corona ($I_c$), Middle Corona ($M_c$) and Outer Corona ($O_c$).

*2) Coronas Division and Nodes Deployment:* Each Corona is divided into four regions as shown in fig. 2. By this division process, we get regions M2, M3, M4 and M5 of middle corona $M_c$, while regions M6, M7, M8 and M9 of outer corona $O_c$. Corona $I_c$, also region M1, is closest to the BS and it is not divided into any further regions. After this, nodes are deployed uniformly over the network field. 20 nodes are deployed in region M1 while 10 nodes are deployed in all the other regions.

### C. Communication Architecture

In our proposed model, data from the sensors reaches BS using multi-hop scheme. It basically consists of 3-Tier communication architecture. In Tier-1, all the non-CH nodes forward their sensed data to their respective CHs. In Tier-2, CHs of the outer corona $O_c$ send their data to the nearest CHs of the middle corona $M_c$. To achieve energy efficiency, CHs of $O_c$ find their distance to the next level CHs and send the data to the nearest CHs. For instance, CH of region $M_6$ finds its distance with the CHs of region $M_2$, $M_5$ and $M_3$, and forwards its data to the minimum distanced CH. In Tier-3, nodes of inner region $I_c$ and CHs of middle region $M_c$ communicate with BS. Fig. 3 depicts this 3-Tier architecture of our implemented work.

### D. Cluster Head Selection

In any clustering protocol, selection of CHs is a very important step. As already stated, our protocol uses static clustering technique, so the number of clusters and CHs remain fixed throughout the network operation. One CH is selected in each region except region 1. Hence, a total of 8 CHs are selected every round.

Table 1
Radio Parameters

| Operations | Energy Dissipated |
|---|---|
| Transmitter/Receiver ETx = ERx | 50 nJ/bit |
| Data Aggregation Energy | 5 nJ/bit/signal |
| εfs (if dtoBS < do) | 10 pJ/bit/4m2 |
| εmp (if dtoBS ≥ do) | 0.0013 pJ/bit/m4 |

Table 2
Protocol Operation Algorithm

| Algorithm | Protocol Operation |
|---|---|
| Step1 ➔ | Do HZ, Deploy Nodes and Initialize Network |
| Step2 ➔ | Compute midpoints of each region with CHs |
| Step3 ➔ | Initialize $N_i$ with nodes of each region $M_i$ |
| Step4 ➔ | Find $N_i$ node's distance with its respective regions midpoint |
| Step5 ➔ | Select CH in descending order of distance |
| Step6 ➔ | **if** CH.$N_i$ belongs to $O_c$ **then**   **if** CH.$N_i$ belongs to $M_6$ **then**     CH.$N_i$.nexthopCH = id.CH of $M_2$   **else** CH.$N_i$ belongs to $M_7$     CH.$N_i$.nexthopCH = id.CH of $M_3$   **else** CH.$N_i$ belongs to $M_8$     CH.$N_i$.nexthopCH = id.CH of $M_4$   **else** CH.$N_i$ belongs to $M_9$     CH.$N_i$.nexthopCH = id.CH of $M_5$   **end if** **end if** |
| Step7 ➔ | **if** the sensed value ≥ Hard Threshold **then**   **if** Hard Threshold is crossed first time **then**     Turn transmitter on and report to CH   **else**     **if** the sensed value ≥ Soft Threshold **then**       Turn transmitter on and report to CH     **end if**   **end if** **end if** |

As mentioned in the communication architecture section that our protocol uses a multi-hop scheme to transmit data to BS, this significantly brings robustness in the network, and energy consumption is controlled to a greater extent. CH is selected on the basis of minimum distance from the center point of the respective zone. In addition, CH selection between region $M_2$ and $M_6$, $M_3$ and $M_7$, $M_4$ and $M_8$, and $M_5$ and $M_9$ is synchronized i.e., if CH in $M_2$ is selected on the right side of the reference point then CH in $M_6$ will also be selected on the right side of the reference point. This results in a smaller distance and helps in avoiding energy and coverage holes.

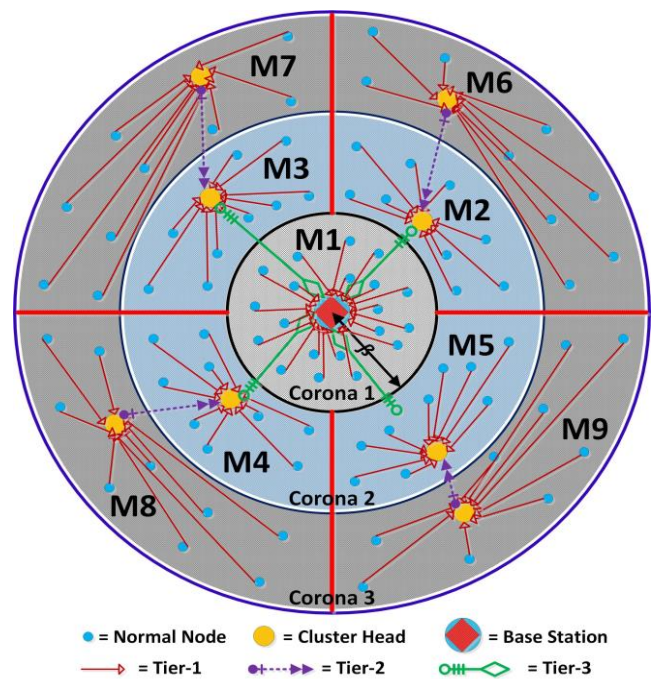

Fig. 3. Communication Architecture

*Correspondance\*: Muzammil Behzad, muzammil.behzad@ieee.org, http://muzammilbehzad.com*

*E. Protocol operation*

In our proposed protocol, reactive as well as heterogeneous nature of the sensor network is taken advantage of, i.e., the nodes with different initial energy levels send their data to BS and Cluster Heads (CHs) only when the sensed attribute crosses a pre-defined threshold, and keep their transmitters off otherwise. We have used the static clustering technique where the network area has been divided into 9 regions. The number of deployed nodes and clusters in these regions will remain static throughout the network lifetime. The regions are named as $M_1, M_2, M_3, ..., M_9$ as shown in fig. 2. Once the network is deployed, the sensors start sensing. The sensors transmit only if the threshold value is crossed. Algorithm 1 describes the stepwise operation of our proposed protocol.

## V. SIMULATIONS, RESULTS AND DISCUSSIONS

In this section, we present and discuss the experimental results of our proposed protocol. The results are compared with the existing traditional protocols on the basis of following metrics: stability period, instability period, network lifetime, number of packets sent, number of packets received and energy consumption.

In our protocol, a total of 100 nodes are deployed randomly in the network area of *100m×100m* which is divided into 9 regions. The position of the base station is at the center of the network field. Initial energy of the nodes of Corona 1 is 0.7*(1+alpha) J where the value of alpha is 1/2, while the initial energies of other nodes is set to 0.7 J. The network is simulated 5 times and average values of the parameters are plotted. Radio parameters used in our simulation are shown in Table 1. Uniform Random Model [13] is also implemented in this scheme to find packets dropped to make this protocol more practical. The probability of the packet drop has been set to 0.3 so as to have more realistic results.

Fig. 4 illustrates the number of packets sent to the base station. It can be seen from the figure that the number of packets sent by DDR [7] is more as compare to that sent in our protocol. The less number of packets sent in our approach is because packets are sent upon crossing the threshold value. Hence, less number of packets will result in less amount of energy consumption and will ultimately prolong the network lifetime. This is depicted in fig. 4 where packets sent by DDR, and the lifetime, end around 3600[th] round while network in our protocol has a lifetime till 6700[th] round.

Fig. 5 shows the packet drop ratio of DDR vs. M-BEHZAD. As can be seen from this figure, the number of packets dropped in M-BEHZAD is smaller as compare to DDR. The reason is threshold sensitive nature of our proposed protocol. Implementation of Uniform Random Model (URM) and the number of packets received per working round has been shown in fig. 6. The results show that the number of packets received on the base station is smaller in the proposed protocol as compare to packets received on BS in DDR.

Fig. 7 demonstrates that the network life time of our approach is 3200 rounds more than DDR [7] and 5300 rounds more than LEACH [1]. The first node die time (FDT) of our proposed protocol is 2200 while that of DDR and LEACH is 1300 and 700 respectively, which proves that the stability period of our protocol is longer. The all node die time (ADT) of proposed protocol is 6700 while that of DDR and LEACH is 3500 and 1400 respectively. The reason behind the longer stability period is the reduced and controlled communication between the nodes to CHs, and CHs to BS. These results validate the efficiency of our protocol. Balanced consumption of energy and uniform random distribution of nodes along with the threshold sensitive nature not only increases network lifetime but it also helps to mitigate energy and coverage holes.

Similarly, in fig. 8, the number of dead nodes per working round is shown. This figure illustrates that the first node in our work dies after 2200[th] round, whereas in DDR and LEACH, after 1300[th] and 700[th] round respectively, which confirms that our protocol is much more efficient than DDR and LEACH in terms of network life time and energy consumption. Also, apart from the longer stability region, the network in our work has a longer instability region which validates our results.

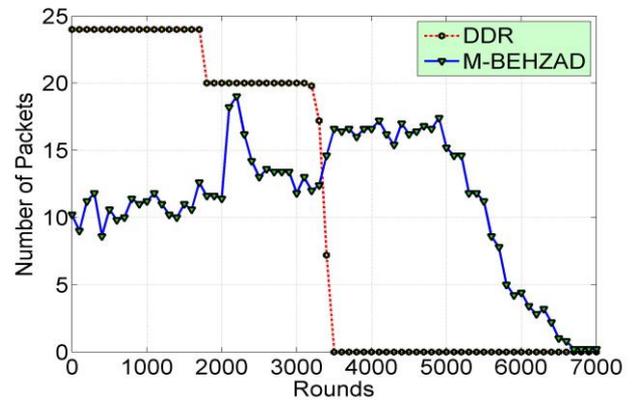

Fig. 4. Packets Sent to BS

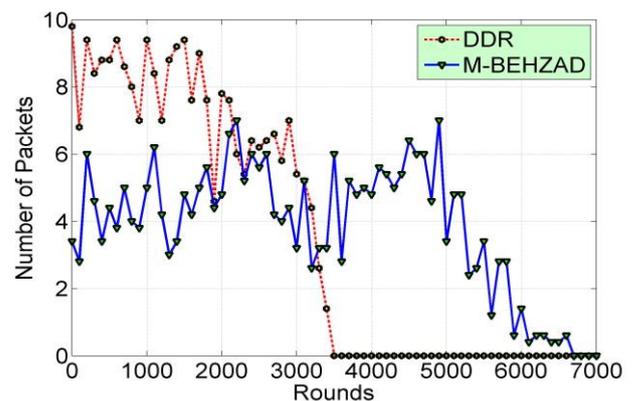

Fig. 5. Packets Dropped Comparison

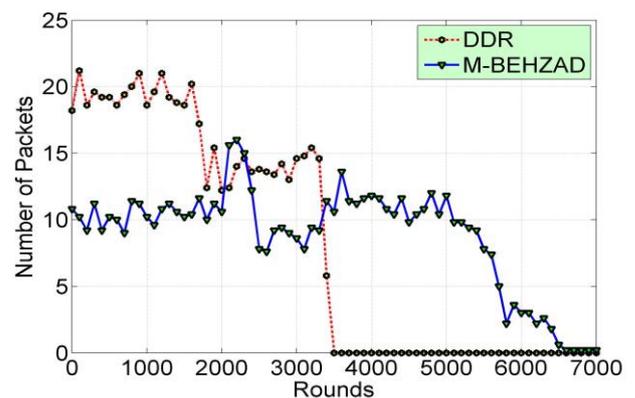

Fig. 6. Packets Received on BS – Uniform Random Model

Fig. 9 shows the propagation delay that each packet experience due to factors like interference, dispersion, refraction, reflection, etc., in non-ideal wireless communication. The delay has been calculated by considering the electromagnetic wavy nature of the signals through which the communication between nodes, CHs and BS take place. The figure shows that delay for each packet is same which is obvious because all packets are communicating through the

*Correspondance\*: Muzammil Behzad, muzammil.behzad@ieee.org, http://muzammilbehzad.com*

electromagnetic signals which move at speed of light when assuming all attenuation factors to be null.

Energy consumption comparison of sensor nodes in DDR and M-BEHZAD has been shown in fig. 10. Both the networks start with same initial network energies. As can be seen, our scheme outperforms DDR in terms of energy consumption as it has used less energy per round. The smaller and controlled energy consumption of our protocol is because of the threshold sensitive nature along with the efficient routing mechanism.

## VI. CONCLUSION

Regarding energy optimization in WSNs, the current research work leads us to examine its existing form. We found that the traditional protocols cannot perform dynamically and are restricted under specific constraints. Our work focuses on effective usage of energy to mitigate both energy and coverage holes in WSNs. M-BEHZAD aims at avoiding frequent communications through threshold dependent scheme along with the heterogeneous nature of sensor network. Sensors in our protocol transmit only if a drastic change in the sensed attribute has occurred. Static clustering with uniform distribution of nodes is used to prolong the network lifetime. Hemisphere Zoning has been introduced for the first time where network field is divided into sliced spherical zones. Thus, CHs are selected based on minimum distance from the reference point of their respective sliced zones. More importantly, 3-Tier communication architecture is implemented to minimize the communication distance.

The performance evaluation of proposed protocol has been carried out and is then compared with LEACH [1] and DDR [7]. Results from simulations confirm that M-BEHZAD outperforms classical routing protocols in terms of stability period, energy consumption and network lifetime. As validated by the results, our approach is 79.1% and 47.8% better than LEACH and DDR, respectively, in terms of stability period.

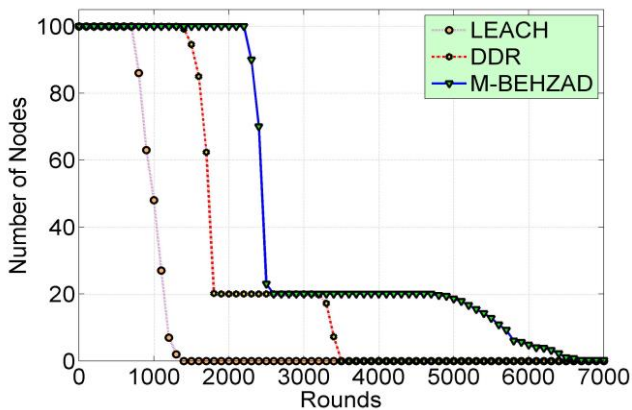
Fig. 7. Allive Nodes

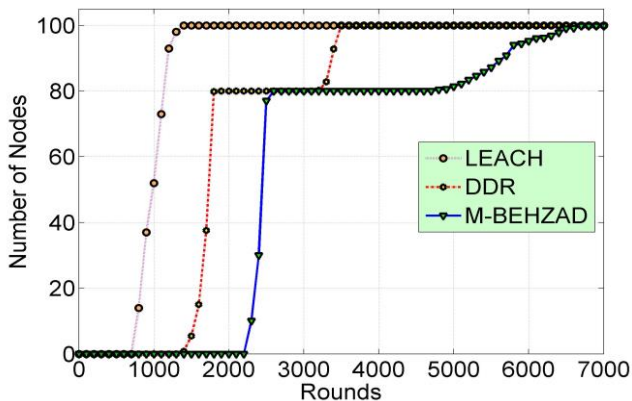
Fig. 8. Stability Period

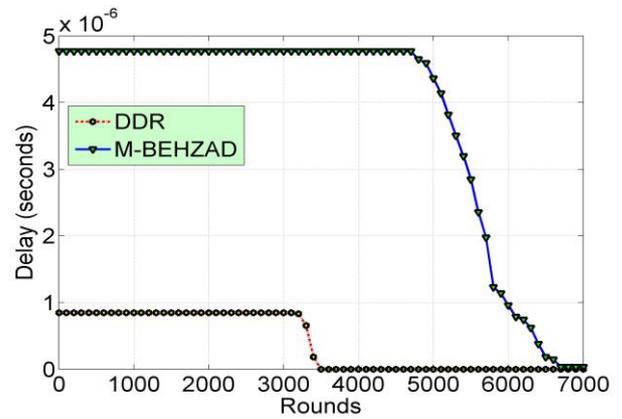
Fig. 9. Propagation Delay

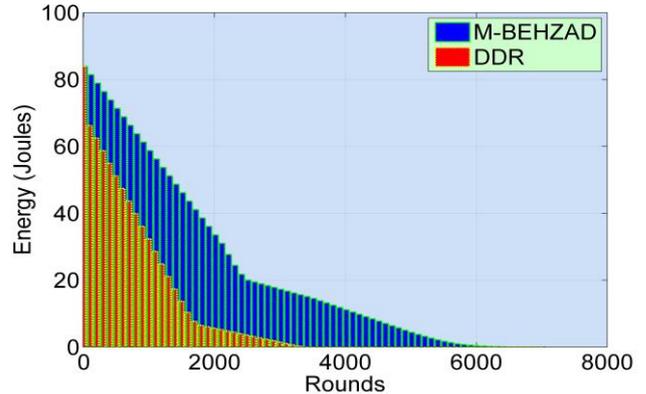
Fig. 10. Energy Consumption

*Correspondance\*: Muzammil Behzad, muzammil.behzad@ieee.org, http://muzammilbehzad.com*

| *Paper Title* | M-BEHZAD: Minimum distance Based Energy efficiency using Hemisphere Zoning with Advanced Divide-and-Rule Scheme for Wireless Sensor Networks | | |
|---|---|---|---|
| *Applicant Name* | Muzammil Behzad | **ID#** | g201402660 |
| *Advisor Name* | Prof. Tareq Y. Al-Naffouri (PhD, Stanford University, U.S.) | | |
| *Field of study* | Wireless Sensor Networks, Wireless Communication | | |

| SUMMARY: (**highlight its originality, importance, and contribution**) |
|---|
| *Routing Protocols* are engaged in a vigorous fashion to ***boost up energy efficiency in WSNs.*** In this paper, ***we propose a novel routing protocol***; Minimum distance Based Energy efficiency using Hemisphere Zoning with Advanced Divide-and-Rule scheme (M-BEHZAD), ***to maximize network lifespan, throughput and stability period*** of the sensors deployed in an un-attended network zone. To accomplish these objectives, ***static clustering technique along with threshold conscious transmissions*** have been used. The robustness of our proposed scheme lies in its ***Cluster Heads (CHs) selection and network field division*** which we are introducing as ***'Hemisphere Zoning (HZ)'***. We have implemented ***3-Tier architecture in order to minimize the communication distance*** which not only ***leads to a better network performance but also significantly reduces the energy and coverage holes, and results in a longer stability period.*** We have also utilized Uniform Random Model (URM) to compute packets dropped to make our scheme a more practical approach. Results from comprehensive simulations using MATLAB validate its applicability. |